\documentclass[twocolumn,pra,aps]{revtex4}
\usepackage{epsfig}

\begin{document}
\title{{\bf Temporary behavior of 'persistent' current in normal rings}}
\author {Zeev~Vager}

\address{Department of Particle  Physics,\\
Weizmann   Institute of Science, 76100 Rehovot, Israel}

\date{ \today}
\begin{abstract}
In equilibrium, the number of conduction electrons in a solid
substance depends on the conformation of the atoms in the
substance. When a magnetic field is applied, it takes time for the
system to come to a new equilibrium with a new conformation.
During such times, current may flow. But, unless superconduction
is involved, the new equilibrium contains no current. It is shown
that the transitory currents are consistent with experiments
involving magnetization of small rings at low temperatures.

\end{abstract}
 \maketitle \bibliographystyle{unsrt}

Persistent current in normal rings was predicted by B\"{u}ttiker,
Imry and Landauer\cite{BIL}. Yet, current transport theories of
conduction fail to explain the observed magnitude of persistent
currents in small normal conducting rings\cite{Webb}. These
theories use perturbation methods on the independent electron
model(e.g. \cite{Yuval}). In the first part of this manuscript
this failure is explained. Then, corrected picture is exploited
for the explanation of the observed persistent current data as
well as a novel insight into transport phenomena. The main point
is illustrated by a simple model.

\emph{\underline{Electrons on a circular ring}}

As a paradigm, a simple model is constructed on a jellium circular
ring. Electrostatically interacting electrons are confined to a
circular ring of homogeneous positive charge which is equal to the
negative charge of the electrons. The static positive charge
exerts an electrostatic potential on the system of electrons which
obey the $N$ body Schr\"{o}dinger equation.

Were the electrons non-interacting, the states of the electron
system had been created from $N$ single electrons of different
angular momentum states, $j^\pi$, for each transversal excitation
where $j$ are half integers. Consistent with the background
potential, the resultant electrons' density would have been
cylindrically symmetric. For non interacting electrons this
density is a sum of uncorrelated, individual electron
cylindrically symmetric densities.

The interaction between electrons introduces a partial rigidity
which manifest through two-body correlation along the azimuthal
angle as well as along transverse dimensions. For very low density
the ground state forms a Wigner crystal which is known to be quite
outside the reach of the independent electron approximation. For
typical conduction electrons density, a remanence of such rigidity
must exist. A rough measure of rigidity is $\Re=(a/a_0)$ (where
$a=k_F^{-1}$ is a typical distance between neighbor electrons and
$a_0$ is the Bohr radius)\cite{Wigner}. This measure is gauged by
the mutual-interaction and the average density of electrons.
Typically, for (3D) conductors, $\Re$ is smaller than it is for a
Wigner crystal by two orders of magnitude. Nevertheless $\Re>0$
and a small finite moment of inertia is formed. Rotational ladders
should appear associated with every internal motion state. For
each state, the density of electrons is cylindrically symmetric
due to the rotational factor, which is a rotating rigid body wave
function.

From this discussion it is clear that the rotational gap is very
large and therefore theories utilizing perturbations on
non-interacting electron picture are so successful. The critical
point is that the validity of the independent electrons
approximation ($\Re=0$) breaks down once a magnetic flux
penetrates the ring. To see that, assume a magnetic flux $\Phi$ in
a form of Aharonov-Bohm (AB) penetrating the ring. Except for
small regions of flux, the very small moment of inertia allows for
a separable hamiltonian and states factorize into rotational and
intrinsic states. The intrinsic states carry no orbital angular
momentum but have quantum numbers of the total spin projection. An
AB flux cannot interact with the spins of electrons, thus, nothing
in the rotating frame interacts with the AB flux. The sum of
orbital angular momenta of individual electrons is the rotational
state angular momentum. Each individual angular momentum is
modified from the 'no-flux' condition to the actual flux condition
by $l\to l+q$, where $q=\Phi/\Phi_0$ and $\Phi_0=hc/e$. In
particular, slowly shifting $q=0\to q=1$, brings a rotational
level from its very high energy location to be the ground
rotational level and takes the former rotational ground state to
very high energy. These huge energy changes with flux are
associated with huge currents. That point was completely neglected
before. In other words, the independent electron theory is not
appropriate for persistent current predictions. The smallest
$\Re>0$ modifies the predictions radically.

Since the rigidity is only a function of the mutual interaction
and the density, it is independent of the actual size of a
homogeneous sample. For example, it is independent of the
circumference of the ring.

\underline{\emph{Symmetry of the deformation}}

Rotational spectra of symmetric top molecules are given
by\cite{Herzberg}
\begin{equation}\label{stm}
    E_K=\frac{\hbar^2 K^2}{2\Im}
\end{equation}
where $K$ is the rotational angular momentum projection along the
axis of symmetry. The allowed $K$ are spaced by $\tilde{N}$ which
characterizes the $\tilde{N}$-fold rotational symmetry in the
rotating frame along the axis. Many times, the allowed $K$ can be
expressed by
\begin{equation}\label{aI}
    K=\tilde{N}j
\end{equation}
where $j$ is either $j_1$ or $j_2$, according to the molecular
identity-of-constituents statistics and
\begin{eqnarray}\label{ii}
% \nonumber to remove numbering (before each equation)
  j_1 &=& \quad\quad\quad\dots,-1,0,1,\cdots \\ \label{iii}
  j_2 &=& \dots,-3/2,-1/2,\,\,1/2,\,\,3/2,\cdots\,.
\end{eqnarray}

The spectrum (\ref{stm}) can be expressed differently with the aid
of $j$:
\begin{equation}\label{stmi}
    E_j=\tilde{N}\frac{\hbar^2 j^2}{2\tilde{\Im}}
    \quad\mbox{where}\quad\tilde{\Im}=\frac{\Im}{\tilde{N}}
\end{equation}
as if $\tilde{N}$ independent and identical particles each with a
small moment of inertia, $\tilde{\Im}$, have identical angular
momenta $j$ which are given by (\ref{ii}) or (\ref{iii}).

This notion of reduced moment of inertia can be applied to the
paradigm above by trying to find the involved $\tilde{N}$ symmetry
in the rotating frame. From the outset assume $N\gg 1$ and that
intrinsic states are not ferromagnetic. Thus, a possible unpaired
spin does not change much the projection of the angular momentum.
Yet, intrinsic states must be antisymmetric. The aim is to find
the highest possible $\tilde{N}$ which is consistent with
interacting electrons, knowing that for non-interacting electrons
theories $\tilde{N}\to\infty$. Clearly, $\tilde{N}$ cannot be
above $N$ - the number of electrons in the system - since any
imaginative intrinsic structure cannot be finer than the number of
the particles involved.  For even $N$, a cyclic permutation of the
electrons indices changes the sign of the state, therefore two
cyclic permutations are equivalent to a unit operator. At maximal
symmetry, this is a rotation by $4\pi/N$. Thus $\tilde{N}=N$ is
the largest possible and $j=j_2$. A cyclic permutation on an odd
$N$ state does not change the state, therefore, at maximal
symmetry, it is equivalent to a rigid rotation of angle $2\pi/N$.
Again $\tilde{N}=N$ is the largest possible with $j=j_1$.

The angular momentum is given by $I=\tilde{N}j$, as if $\tilde{N}$
identical boson particles are at the same quantum number. This is
just the property of the rotational factor of the state. The
intrinsic state must stay antisymmetric.

In the independent electron model an electron on a circle with
radius $R$ at the Fermi level has an angular momentum of $\pm|k_F
R|$. Thus, for each channel there are $2E_F m R^2/\hbar^2=(k_F
R)^2$ electrons. Using this number for $\tilde{N}$ in
(\ref{stmi}), the rotational constant would have been $E_F$ if
$\tilde{\Im}=mR^2$. But that would have meant that the rigidity is
extremely high. Therefore, define $E_>=E_F(m/m^*)$ for the
rotational constant where $m^*\ll m$ is a very small effective
mass representing the small rigidity of electrons in the rotating
frame.

When many channels are involved, the symmetry $\tilde{N}$ is
limited to the number of electrons along one channel. In this
approximation $\tilde{N}=2\pi k_F R=2\pi R/a$. The rotational
constant is still very high. Therefore, the separability of
rotations and intrinsic structure is valid for conductivity
experiments where temperatures are much lower than the rotational
constant.

The parameters $E_F$ and $k_F$ are just representatives of the
average electrons density. In this interpretation the rotational
states behave like a symmetric top with $\tilde{N}=2\pi R/a$ -
fold symmetry where the rotational energies are given by
\begin{equation}\label{RE}
    E_j=E_> j^2\gg E_F j^2\,.
\end{equation}
As a preparation for later discussion, notice that in this
paradigm there is a huge energetic preference for odd $N$ with
$j=0$.

\underline{\emph{Orbital magnetism}}

For any system of electrons, given the total orbital angular
momentum, $I$, the orbital magnetic moment is
$I\mu_B=\tilde{N}j\mu_B$. This changes dramatically when a
magnetic flux, $\Phi$, passes through the ring. Normalizing the AB
flux to a parameter $q=\Phi/\Phi_0$ where $\Phi_0=hc/e$, the
magnetic moment is given by
\begin{equation}\label{mm}
    M=(\frac{2\pi R}{a}) (j+q)\mu_B\,.
\end{equation}
The behavior near zero flux is very different for odd and even $N$
since from equations (\ref{ii},\ref{iii}) $j=j_1$ and $j=j_2$
respectively. There is an involved current, $M/\pi R^2$, with any
such magnetic moment. In this paradigm, changing the magnetic flux
by $\delta q$ induces current $\delta M /\pi R^2= 2\delta
q\mu_B/Ra=\delta q ev_F/cR$. It has nothing to do with the Fermi
distribution. Its source is a global rotation of the intrinsic
structures.

If the operator of interest is only the orbital magnetic moment, a
truncated density matrix can be used where all degrees of freedom,
except for rotational, are traced out. The advantage of it is that
the number of needed states of $N$ electrons reduces dramatically
to the number of needed (perturbed) rotational states. This is
followed next.

\underline{\emph{The paradigm as a representation}}

Two important steps are carried out towards approaching real
physical situations. The first is breaking away from cylindrical
symmetry. This is dealt with in this section. The other is the
issue of dynamics of atoms locations and environment.

The cylindrical symmetry is lost when the positive charge is
fragmented to fixed locations of ions. The neutralizing electrons
are considered as $N$ conduction electrons. In this unique
situation there is no mechanism for dissipation. Therefore, even
without going to details, the ring can be considered as a type of
a Josephson ring. Nevertheless, details follow.

In a fixed conformation (locations of ions), the system in the
loop has a discrete set of states. Therefore, for any given flux,
$q$, each state $|i,q\rangle$ is a superposition of the paradigm
unperturbed states at the same flux. The latter have rotational
factors depending on $X$.
\begin{equation}\label{iq}
    |i,q\rangle=\sum_{j,k} a_{n,k}^i(q)\Psi_j(X)\phi_k
\end{equation}
where $\Psi_j(X)$ are the rotational states and $\phi$ are
intrinsic functions, both  flux independent\cite{FN1}. The one
dimension $X$ represents either an angle of rotation or any
related single coordinate along the loop, such as the center of
mass of the electrons along the ring normalized to the
circumference. Each such state has an associated magnetization (in
units of Bohr magnetons)
\begin{eqnarray}
\nonumber %to remove numbering (before each equation)
  \langle i|\textbf{L}|i\rangle=
    \sum_{j',j,k',k}a^{i*}_{j',k'}a^i_{j,k}
    \langle \Psi_{j'}|\textbf{L}|\Psi_{j}\rangle
    \delta_{k',k}\delta_{j',j} \\
  \equiv \sum_n b^i_j(q) (\frac{2\pi R}{a}) (j+q)
\end{eqnarray}
where $b^i_j\geq 0$ and $(\frac{2\pi R}{a}) (j+q)$ is the
magnetization of the $j$-th rotational state.

The magnetization of the ensemble is
\begin{eqnarray}\label{ens-mag}
\nonumber
    M(q)=\sum_i w_i(q) \langle i|\textbf{L}|i\rangle
    &=&\sum_i w_i(q)\sum_j b^i_j (\frac{2\pi R}{a}) (j+q) \\
    &=&(\frac{2\pi R}{a}) \sum_j w'_j(q)(j+q)
\end{eqnarray}
where $w'_j=\sum_i w_i b^i_j$ and $\sum_j w'_j(q)=1$. The result
proves that the magnetization of the ensemble is given by a
weighted sum of the magnetization of the rotational states.

Define $j_0$ as a number from (\ref{ii}) for odd $N$ or a number
from (\ref{iii}) for even $N$ for which $|j_0+q|\leq 1/2$. The
magnetization of the rotational ground state is the sharp
saw-tooth function
\begin{equation}\label{M_0}
    M_0(q)=(\frac{2\pi R}{a}) (j_0+q)\quad ;\quad |j_0+q|\leq 1/2
\end{equation}
and the magnetization of the perturbed ground state is a blunt
saw-tooth function
\begin{equation}\label{Tmag}
    M(q)=T(q)M_0(q)
\end{equation}
where $T(q)$ is a non-negative real unit-less number, defined by
equations (\ref{ens-mag}) and (\ref{M_0}). It is independent of
$R/a$. Similarly, for the persistent current
\begin{equation}\label{Tcurr}
    I(q)=T(q)I_0(q)
\end{equation}
and  $T(q)$ can be identified as the transmission of the current
in the ring which is penetrated by a flux $q$. The independency of
$T(q)$ from $N$ is important for extensions where $N$ may vary.

Similar equations to (\ref{Tmag}) and (\ref{Tcurr}) can be found
for Josephson rings where the interpretation of $T(q)$ is the
transmission of the current in the ring which is penetrated by a
flux $q$. This transmission is not one since the Josephson ring is
not a continuous superconductor but contains barriers. As was
shown by Bloch\cite{Bloch}, this follows from general physical
principles without the need to have a detailed model for
superconduction. In an analogue way the fixed conformation ring
can be visualized by domains of perfect conductors surrounded by
barriers. In the fixed conformation picture, the current is a
coherent phenomenon in the same fashion as in a Josephson ring
with penetrating magnetic flux.

\emph{\underline{The non-dissipative resistance}}

If a fixed conformation ring or a Josephson ring is opened and
connected in series to many identical replica, the system remains
lossless. But an infinitely long chain is like a transmission line
and has its characteristic real impedance - $Z_0$. Since this
resistance is independent of flux, it must be given by the
averaged flux transmission (in other words, the boundary condition
at infinity is undefined)
\begin{equation}\label{afT}
    {\cal T}=\langle T(q) \rangle\,.
\end{equation}
Define
\begin{equation}\label{ndr}
    Z_0=\frac{\hbar\pi}{e^2}\frac{1-{\cal T}}{\cal T}\,.
\end{equation}
When $Z_0\ll\frac{\hbar\pi}{e^2}$ then $\cal T$ is very close to
one. The function $T(q)$ is known to have zeros, therefore, in
almost all other flux values in must be even closer to one than
$\cal T$. From equation (\ref{Tmag}), the factor $T(q)$ limits the
maximum orbital magnetization. Thus, the condition
$Z_0\ll\frac{\hbar\pi}{e^2}$ assures an almost maximal
magnetization.

This concludes the discussion for the strict case of conduction
electrons with fixed conformation of the ion-donors. The
characteristic impedance controls the maximum magnetization like
in a Josephson junction. But, differently from it, the number of
electrons is fixed and, by equation (\ref{M_0}),  the maximum
possible magnetization is $(\frac{\pi R}{a})\mu_B$\cite{current}.

It might be argued that the reduced density matrix $\rho(X',X)$ in
the fixed conformation approximation has an off diagonal long
range order\cite{Yang}. Therefore, the rigidity is connected to a
long range two-electron correlation which is independent of size.
This approximate condensation is different from superconduction
because $N$ is fixed.

\underline{\emph{Relaxation of the fixed conformation condition}}

A striking puzzle of the fixed conformation condition is that at
many flux values, odd and even $N$ have very different rotational
energies with a gap reaching a rotational constant. This puzzle
remains even for the perturbed rotational states when
$Z_0<\frac{\hbar\pi}{e^2}$ where the above gap is somewhat
smaller. Another puzzle seems to be the traditional result that
persistent current can be sustained in a 'normal' conductor having
non-zero resistance. The resolution of these puzzles is that at
equilibrium there are many fixed conformations sustaining an even
number conduction electrons and many other sustaining a nearby odd
number and anywhere in-between. Therefore, in equilibrium, no
persistent current flows in normal conductors. In general, this
multiplicity of conformations makes the passage from one
conformation to the other an irreversible path. The relaxation
process between different flux conditions takes time which is
strongly dependent on temperature. If measurements of
magnetization are carried out during times shorter then the
relaxation time, the system is driven to a non-equilibrium state
which sustains current. Large orbital magnetization is associated
with large rotational energy for fixed $N$. Eventually, this will
relax into $N\to N'$ with no orbital magnetization and zero
rotational energy. Thus, the fixed conformation approximation is
valid for times shorter than a relaxation time.

The temperature behavior of resistivity in non-magnetic normal
conductors is a text-book material\cite{Ziman}. It is given by the
Bloch-Gr\"{u}neisen formula as $\rho=\rho_0+\rho_{el-ph}(T/T_D)$
where $\rho_0$ is the residual resistivity due to defect
scattering and is essentially temperature independent. The
temperature dependent part of the resistivity, $\rho_{el-ph}$,
arises from electron-phonon interaction and $T_D$ is the Debye
temperature. When the temperature is low enough, only residual
resistivity remains. This is usually referred to as 'the elastic
regime'.

This picture can be restated in the framework of this article as
follows. The positions of the positive 'background' ions
(conformations) is usually described by displacements from
equilibrium positions. This very general method leads to a phonon
spectrum. A strict application of the method assumes that at zero
temperature there is a ground state conformation. Regular solids
have very many ground state conformations and almost all of them
are disordered. Moreover, the passage between almost any two such
conformations involves high barriers which are not simply
described in terms of a phonon theory. When magnetic field enters,
a ground state of a given conformation may be an excited state of
another conformation. Thus, quasi-equilibrium states of the
electron-ion system may survive for much longer time than
predicted by an application of a Debye based theory.

Therefore, for many mesoscopic systems at low enough temperature
the relaxation time is long and for rings, leaving ample time for
physical measurements where the residual resistivity can be
ignored. The difference between the open loop residual resistance
and the closed loop residual resistance is analogue to the
difference in transmission, given by equation (\ref{afT}). While
$T(q)$ describes a totally coherent transmission, its average,
$\cal T$ describes losses to the contacts due to random phases.
Thus, the characteristic impedance of a loop is the residual
resistance in an open loop.

\underline{\emph{A transmission line analogue}}

Classically, a transmission line (TL) is characterized by the
inductance and capacitance per unit length, $\cal L$ and $\cal C$
as well as ohmic resistance per unit length $\cal R$. The
characteristic impedance is given by $Z_0=\sqrt{{\cal L}/{\cal
C}}$. While $Z_0$ has almost no temperature dependence, for normal
conductors the value of $R$ decreases with temperature until some
minimum. As above, consider a hypothetical TL built from replica
of sections of length $L$ connected in series. At the elastic
regime ${\cal R}=0$ for any measurement performed in times shorter
than the relaxation time. As shown above, this result is quantal.

The unit length corresponds to the circumference of a ring. If a
step function voltage, $V$, is applied to the TL, or equivalently
an EMF is induced in the ring, it will draw a current of $V/Z_0$
and the disturbance travels along the TL, or, circling the ring.
Eventually, at times comparable to the relaxation time, the
current decays by a change in conformations process. In analogy to
an open Josephson junction replica TL, ${\cal C}^{-1}$ represents
the gap-barriers and increases linearly with the length and so is
$\cal L$. Therefore, $Z_0$ is proportional to the circumference as
it should for residual resistance.

\underline{\emph{Conclusions}}

At the elastic regime, normal conductors have a residual
resistance which is associated with dissipation. This dissipation
temporarily disappears when connected into a loop. The residual
resistance becomes the non-dissipative impedance which determine
the maximal magnetization. This coherent metastable state depends
on 'initial conditions' which selects appropriate metastable
conformations. For residual resistance lower than the universal
resistance, almost full magnetization is expected and this is
consistent with observations.

The physics advocated here for transport phenomena is different
than customary. In particular, it indicates the importance of
collective transport, similar to superconduction.

Illuminating discussions with Profs. Yuval Oreg, David Mukamel,
Zvi Lipkin and Ron Naaman are acknowledged.


\begin{thebibliography}{99}
\bibitem{Wigner} A dimensionless analysis for the condition of a
Wigner lattice is as follows. A ratio, $\Re$, between the average
potential energy of an electron and its kinetic energy is defined
as rigidity. This ratio should be bigger than some constant
(derived by numerical calculations) for maintaining a crystal
structure.
\begin{eqnarray}\label{oreg}
    \nonumber
    \frac{\mbox{pot.}}{\mbox{kin.}}\propto\frac{e^2 k_F}{\hbar^2
    k_F^2/2m}\propto(k_F a_0)^{-1}\equiv\Re
\end{eqnarray}
where $a_0$ is the Bohr radius.
\bibitem{FN1} Both the rotational and the intrinsic
states are independent of the magnetic flux. The former is due to
the cylindrically symmetric paradigm system. The intinsic state
independency from flux is due to its zero total orbital momentum.
\bibitem{current} The maximum current, in terms of $v_F$ is
$I_0=ev_F/2R$.
\end{thebibliography}
\end{document}